\documentclass[12pt]{article}
\pdfoutput=1
\usepackage[]{hyperref}
\usepackage{xcolor}
\usepackage[]{putex}
\usepackage[utf8]{inputenc}

\usepackage{amsmath,amsthm,amssymb}
\usepackage{tikz}
\usepackage{tikz-cd}
\usetikzlibrary{decorations.markings}
\usetikzlibrary{calc}

\makeatletter
\DeclareRobustCommand*{\bfseries}{%
  \not@math@alphabet\bfseries\mathbf
  \fontseries\bfdefault\selectfont
  \boldmath
}
\makeatother
\hypersetup{
  colorlinks=true,
  linkcolor=blue,
  citecolor=green!60!black,
  urlcolor=cyan!70!black,
  linktoc=all
  }

\input{glyphtounicode}
\def\musepic#1{\vcenter{\hbox{\usebox{#1}}}}
\newmuskip\pFqmuskip

\newcommand*\pFq[6][8]{%
  \begingroup 
  \pFqmuskip=#1mu\relax
  \mathchardef\normalcomma=\mathcode`,
  \mathcode`\,=\string"8000
  \begingroup\lccode`\~=`\,
  \lowercase{\endgroup\let~}\pFqcomma
  {}_{#2}F_{#3}{\left[\genfrac..{0pt}{}{#4}{#5};#6\right]}%
  \endgroup
}
\newcommand{\pFqcomma}{{\normalcomma}\mskip\pFqmuskip}

\newsavebox{\figcalWpadic}
\savebox{\figcalWpadic}{%
\begin{tikzpicture}[scale=.9]
    \draw[thick, color=black!30] (0,0) circle[radius=2];
    \coordinate (x1) at (200:2);
    \coordinate (x2) at (160:2);
    \coordinate (x4) at (20:2);
    \coordinate (x5) at (-20:2);
    \coordinate (x3) at (90:2);
    \coordinate (y) at (180:1.3);
    \coordinate (yp) at (0:1.3);
	\draw[thick, color=red, dashed] (x2) to [out =  -15, in = 15, looseness = 1.5] (x1);
	\draw[thick, color=red, dashed] (x4) to [out =  195, in = 165, looseness = 1.5] (x5);
	\draw[thick] (x1)--(y)--(x2);
	\draw[thick] (y)--(x3)--(yp);
	\draw[thick] (x4)--(yp)--(x5);
	\draw[thick] (y)--(yp);
	\draw (x1) node[anchor=north east] {${\cal O}_1$};
	\draw (x2) node[anchor=south east] {${\cal O}_2$};
	\draw (x3) node[anchor=south] {${\cal O}_3$};
	\draw (x4) node[anchor=south west] {${\cal O}_4$};
	\draw (x5) node[anchor=north west] {${\cal O}_5$};
	\draw[color=blue]  ($(y)!0.5!(x3)$) node[anchor=south east] {$\Delta_{3a,b}$};
	\draw[color=blue]  ($(yp)!0.5!(x3)$) node[anchor=south west] {$\Delta_{3b,a}$};
	\draw[color=blue]  ($(y)!0.5!(yp)$) node[anchor=north] {$\Delta_{ab,3}$};
\end{tikzpicture}}
\newsavebox{\figcalW}
\savebox{\figcalW}{%
\begin{tikzpicture}[scale=1]
    \draw[thick, color=black!30] (0,0) circle[radius=2];
    \coordinate (x1) at (200:2);
    \coordinate (x2) at (160:2);
    \coordinate (x4) at (20:2);
    \coordinate (x5) at (-20:2);
    \coordinate (x3) at (90:2);
    \coordinate (y) at (180:1.3);
    \coordinate (yp) at (0:1.3);
	\draw[thick, color=red, dashed] (x2) to [out =  -15, in = 15, looseness = 1.5] (x1);
	\draw[thick, color=red, dashed] (x4) to [out =  195, in = 165, looseness = 1.5] (x5);
	\draw[thick] (x1)--(y)--(x2);
	\draw[thick] (y)--(x3)--(yp);
	\draw[thick] (x4)--(yp)--(x5);
	\draw[thick, dotted] (y)--(yp);
	\draw (x1) node[anchor=north east] {${\cal O}_1$};
	\draw (x2) node[anchor=south east] {${\cal O}_2$};
	\draw (x3) node[anchor=south] {${\cal O}_3$};
	\draw (x4) node[anchor=south west] {${\cal O}_4$};
	\draw (x5) node[anchor=north west] {${\cal O}_5$};
	\draw[color=blue]  ($(y)!0.6!(x3)$) node[anchor=south east, text width=1.5cm, align=center] {$\Delta_{3a,b}$\\$-k_1+k_2$};
	\draw[color=blue]  ($(yp)!0.6!(x3)$) node[anchor=south west, text width=1.5cm, align=center] {$\Delta_{3b,a}$\\$+k_1-k_2$};
	\draw[color=blue]  ($(y)!0.5!(yp)$) node[anchor=north, text width=3cm, align=center] {$\Delta_{ab,3}+k_1+k_2$};
\end{tikzpicture}}
\newsavebox{\figcalWalt}
\savebox{\figcalWalt}{%
\begin{tikzpicture}[scale=1.2]
    \draw[thick, color=black!30] (0,0) circle[radius=2];
    \coordinate (x1) at (200:2);
    \coordinate (x2) at (160:2);
    \coordinate (x4) at (20:2);
    \coordinate (x5) at (-20:2);
    \coordinate (x3) at (90:2);
    \coordinate (y) at (180:1.3);
    \coordinate (yp) at (0:1.3);
	\draw[thick, color=red, dashed] (x2) to [out =  -15, in = 15, looseness = 1.5] (x1);
	\draw[thick, color=red, dashed] (x4) to [out =  195, in = 165, looseness = 1.5] (x5);
	\draw[thick] (x1)--(y)--(x2);
	\draw[thick] (y)--(x3)--(yp);
	\draw[thick] (x4)--(yp)--(x5);
	\draw[thick] (y)--(yp);
	\draw (x1) node[anchor=north east] {${\cal O}_1$};
	\draw (x2) node[anchor=south east] {${\cal O}_2$};
	\draw (x3) node[anchor=south] {${\cal O}_3$};
	\draw (x4) node[anchor=south west] {${\cal O}_4$};
	\draw (x5) node[anchor=north west] {${\cal O}_5$};
	\draw[color=blue]  ($(y)!0.6!(x3)$) node[anchor=south east, text width=1.5cm, align=center] {$\Delta_{3a,b}$\\$-k_1+k_2$};
	\draw[color=blue]  ($(yp)!0.6!(x3)$) node[anchor=south west, text width=1.5cm, align=center] {$\Delta_{3b,a}$\\$+k_1-k_2$};
	\draw[color=blue]  ($(y)!0.5!(yp)$) node[anchor=north, text width=3.2cm, align=center] {$\Delta_{ab,3}+k_1+k_2+2k_3$};
\end{tikzpicture}}
\newsavebox{\figcalWGdef}
\savebox{\figcalWGdef}{%
\begin{tikzpicture}[scale=.8]
    \draw[thick, color=black!30] (0,0) circle[radius=2];
    \coordinate (x1) at (200:2);
    \coordinate (x2) at (160:2);
    \coordinate (x4) at (20:2);
    \coordinate (x5) at (-20:2);
    \coordinate (x3) at (90:2);
    \coordinate (y) at (180:1.3);
    \coordinate (yp) at (0:1.3);
	\draw[thick, color=red, dashed] (x2) to [out =  -15, in = 15, looseness = 1.5] (x1);
	\draw[thick, color=red, dashed] (x4) to [out =  195, in = 165, looseness = 1.5] (x5);
	\draw[thick] (x1)--(y)--(x2);
	\draw[thick] (y)--(x3)--(yp);
	\draw[thick] (x4)--(yp)--(x5);
	\draw[thick] (y)--(yp);
	\draw (x1) node[anchor=north east] {${\cal O}_1$};
	\draw (x2) node[anchor=south east] {${\cal O}_2$};
	\draw (x3) node[anchor=south] {${\cal O}_3$};
	\draw (x4) node[anchor=south west] {${\cal O}_4$};
	\draw (x5) node[anchor=north west] {${\cal O}_5$};
	\draw[color=blue]  ($(y)!0.5!(x3)$) node[anchor=east] {$\Delta_{L}$};
	\draw[color=blue]  ($(yp)!0.5!(x3)$) node[anchor=west] {$\Delta_{R}$};
	\draw[color=blue]  ($(y)!0.5!(yp)$) node[anchor=north] {$\Delta_{C}$};
\end{tikzpicture}}
\newsavebox{\figcalWdef}
\savebox{\figcalWdef}{%
\begin{tikzpicture}[scale=.8]
    \draw[thick, color=black!30] (0,0) circle[radius=2];
    \coordinate (x1) at (200:2);
    \coordinate (x2) at (160:2);
    \coordinate (x4) at (20:2);
    \coordinate (x5) at (-20:2);
    \coordinate (x3) at (90:2);
    \coordinate (y) at (180:1.3);
    \coordinate (yp) at (0:1.3);
	\draw[thick, color=red, dashed] (x2) to [out =  -15, in = 15, looseness = 1.5] (x1);
	\draw[thick, color=red, dashed] (x4) to [out =  195, in = 165, looseness = 1.5] (x5);
	\draw[thick] (x1)--(y)--(x2);
	\draw[thick] (y)--(x3)--(yp);
	\draw[thick] (x4)--(yp)--(x5);
	\draw[thick, dotted] (y)--(yp);
	\draw (x1) node[anchor=north east] {${\cal O}_1$};
	\draw (x2) node[anchor=south east] {${\cal O}_2$};
	\draw (x3) node[anchor=south] {${\cal O}_3$};
	\draw (x4) node[anchor=south west] {${\cal O}_4$};
	\draw (x5) node[anchor=north west] {${\cal O}_5$};
	\draw[color=blue]  ($(y)!0.5!(x3)$) node[anchor=east] {$\Delta_{L}$};
	\draw[color=blue]  ($(yp)!0.5!(x3)$) node[anchor=west] {$\Delta_{R}$};
	\draw[color=blue]  ($(y)!0.5!(yp)$) node[anchor=north] {$\Delta_{C}$};
\end{tikzpicture}}
\newsavebox{\figcalWdefC}
\savebox{\figcalWdefC}{%
\begin{tikzpicture}[scale=.8]
    \draw[thick, color=black!30] (0,0) circle[radius=2];
    \coordinate (x1) at (200:2);
    \coordinate (x2) at (160:2);
    \coordinate (x4) at (20:2);
    \coordinate (x5) at (-20:2);
    \coordinate (x3) at (90:2);
    \coordinate (y) at (180:1.3);
    \coordinate (yp) at (0:1.3);
	\draw[thick, color=red, dashed] (x2) to [out =  -15, in = 15, looseness = 1.5] (x1);
	\draw[thick, color=red, dashed] (x4) to [out =  195, in = 165, looseness = 1.5] (x5);
	\draw[thick] (x1)--(y)--(x2);
	\draw[thick] (y)--(x3)--(yp);
	\draw[thick] (x4)--(yp)--(x5);
	\draw[thick, dotted] (y)--(yp);
	\draw (x1) node[anchor=north east] {${\cal O}_1$};
	\draw (x2) node[anchor=south east] {${\cal O}_2$};
	\draw (x3) node[anchor=south] {${\cal O}_3$};
	\draw (x4) node[anchor=south west] {${\cal O}_4$};
	\draw (x5) node[anchor=north west] {${\cal O}_5$};
	\draw[color=blue]  ($(y)!0.5!(x3)$) node[anchor=east] {$\Delta_{L}$};
	\draw[color=blue]  ($(yp)!0.5!(x3)$) node[anchor=west] {$\Delta_{R}$};
	\draw[color=blue]  ($(y)!0.5!(yp)$) node[anchor=north] {$\Delta_{C}+2$};
\end{tikzpicture}}
\newsavebox{\figcalWdefLCR}
\savebox{\figcalWdefLCR}{%
\begin{tikzpicture}[scale=.8]
    \draw[thick, color=black!30] (0,0) circle[radius=2];
    \coordinate (x1) at (200:2);
    \coordinate (x2) at (160:2);
    \coordinate (x4) at (20:2);
    \coordinate (x5) at (-20:2);
    \coordinate (x3) at (90:2);
    \coordinate (y) at (180:1.3);
    \coordinate (yp) at (0:1.3);
	\draw[thick, color=red, dashed] (x2) to [out =  -15, in = 15, looseness = 1.5] (x1);
	\draw[thick, color=red, dashed] (x4) to [out =  195, in = 165, looseness = 1.5] (x5);
	\draw[thick] (x1)--(y)--(x2);
	\draw[thick] (y)--(x3)--(yp);
	\draw[thick] (x4)--(yp)--(x5);
	\draw[thick, dotted] (y)--(yp);
	\draw (x1) node[anchor=north east] {${\cal O}_1$};
	\draw (x2) node[anchor=south east] {${\cal O}_2$};
	\draw (x3) node[anchor=south] {${\cal O}_3$};
	\draw (x4) node[anchor=south west] {${\cal O}_4$};
	\draw (x5) node[anchor=north west] {${\cal O}_5$};
	\draw[color=blue]  ($(y)!0.75!(x3)$) node[anchor=east] {$\Delta_{L}-1$};
	\draw[color=blue]  ($(yp)!0.75!(x3)$) node[anchor=west] {$\Delta_{R}+1$};
	\draw[color=blue]  ($(y)!0.5!(yp)$) node[anchor=north] {$\Delta_{C}+1$};
\end{tikzpicture}}
\newsavebox{\figcalWfour}
\savebox{\figcalWfour}{%
\begin{tikzpicture}[scale=.8]
    \draw[thick, color=black!30] (0,0) circle[radius=2];
    \coordinate (x1) at (200:2);
    \coordinate (x2) at (160:2);
    \coordinate (x4) at (20:2);
    \coordinate (x5) at (-20:2);
    \coordinate (x3) at (90:2);
    \coordinate (y) at (180:1.3);
    \coordinate (yp) at (0:1.3);
	\draw[thick, color=red, dashed] (x2) to [out =  -15, in = 15, looseness = 1.5] (x1);
	\draw[thick, color=red, dashed] (x4) to [out =  195, in = 165, looseness = 1.5] (x5);
	\draw[thick] (x1)--(y)--(x2);
	\draw[thick] (x4)--(yp)--(x5);
	\draw[thick] (y)--(yp);
	\draw (x1) node[anchor=north east] {${\cal O}_1(x_1)$};
	\draw (x2) node[anchor=south east] {${\cal O}_2(x_2)$};
	\draw (x4) node[anchor=south west] {${\cal O}_3(x_3)$};
	\draw (x5) node[anchor=north west] {${\cal O}_4(x_4)$};
	\draw  ($(y)!0.5!(yp)$) node[anchor=north, color=blue] {$\Delta$};
\end{tikzpicture}}
\newsavebox{\figcalWfourxi}
\savebox{\figcalWfourxi}{%
\begin{tikzpicture}[scale=.8]
    \draw[thick, color=black!30] (0,0) circle[radius=2];
    \coordinate (x1) at (200:2);
    \coordinate (x2) at (160:2);
    \coordinate (x4) at (20:2);
    \coordinate (x5) at (-20:2);
    \coordinate (x3) at (90:2);
    \coordinate (y) at (180:1.3);
    \coordinate (yp) at (0:1.3);
	\draw[thick, color=red, dashed] (x2) to [out =  -15, in = 15, looseness = 1.5] (x1);
	\draw[thick, color=red, dashed] (x4) to [out =  195, in = 165, looseness = 1.5] (x5);
	\draw[thick] (x1)--(y)--(x2);
	\draw[thick] (x4)--(yp)--(x5);
	\draw[thick, dotted] (y)--(yp);
	\draw (x1) node[anchor=north east] {${\cal O}_1$};
	\draw (x2) node[anchor=south east] {${\cal O}_2$};
	\draw (x4) node[anchor=south west] {${\cal O}_4$};
	\draw (x5) node[anchor=north west] {${\cal O}_5$};
	\draw[color=blue]  ($(y)!0.5!(yp)$) node[anchor=north] {$\Delta_{a}+2k_1$};
\end{tikzpicture}}
\newsavebox{\figOPElim}
\savebox{\figOPElim}{%
\begin{tikzpicture}[scale=.8]
    \draw[thick, color=black!30] (0,0) circle[radius=2];
    \coordinate (x1) at (200:2);
    \coordinate (x2) at (160:2);
    \coordinate (x4) at (20:2);
    \coordinate (x5) at (-20:2);
    \coordinate (x3) at (90:2);
    \coordinate (y) at (180:1.3);
    \coordinate (yp) at (0:1.3);
	\draw[thick, color=red, dashed] (x2) to [out =  -15, in = 15, looseness = 1.5] (x1);
	\draw[thick] (x1)--(y)--(x2);
	\draw[thick] (y)--(x4);
	\draw[thick] (y)--(x5);
	\draw (x1) node[anchor=north east] {${\cal O}_1(x_1)$};
	\draw (x2) node[anchor=south east] {${\cal O}_2(x_2)$};
	\draw (x4) node[anchor=south west] {${\cal O}_3(x_3)$};
	\draw (x5) node[anchor=north west] {${\cal O}_4(x_4)$};
	\draw[color=blue]  ($(y)!0.5!(x4)$) node[anchor=south, text width=1.6cm, align=center] {$\Delta_{3a,b} +k_2$};
	\draw[color=blue]  ($(y)!0.5!(x5)$) node[anchor=north, text width=1.6cm, align=center] {$\Delta_{ab,3} +k_2$};
\end{tikzpicture}}
\newsavebox{\figcalWfourOPE}
\savebox{\figcalWfourOPE}{%
\begin{tikzpicture}[scale=.8]
    \draw[thick, color=black!30] (0,0) circle[radius=2];
    \coordinate (x1) at (200:2);
    \coordinate (x2) at (160:2);
    \coordinate (x4) at (20:2);
    \coordinate (x5) at (-20:2);
    \coordinate (x3) at (90:2);
    \coordinate (y) at (180:1.3);
    \coordinate (yp) at (0:1.3);
	\draw[thick, color=red, dashed] (x2) to [out =  -15, in = 15, looseness = 1.5] (x1);
	\draw[thick, color=red, dashed] (x4) to [out =  195, in = 165, looseness = 1.5] (x5);
	\draw[thick] (x1)--(y)--(x2);
	\draw[thick] (x4)--(yp)--(x5);
	\draw[thick] (y)--(yp);
	\draw (x1) node[anchor=north east] {${\cal O}_1(x_1)$};
	\draw (x2) node[anchor=south east] {${\cal O}_2(x_2)$};
	\draw (x4) node[anchor=south west] {${\cal O}_3(x_3)$};
	\draw (x5) node[anchor=north west] {${\cal O}_b(x_4)$};
	\draw  ($(y)!0.5!(yp)$) node[anchor=north] {$\Delta_{a}$};
\end{tikzpicture}}
\newsavebox{\figFourPtA}
\savebox{\figFourPtA}{%
\begin{tikzpicture}[scale=.8]
    \tikzstyle{vint}=[draw,scale=0.55,color=teal,fill=green,circle]
    \draw[thick, color=black!30] (0,0) circle[radius=2];
    \coordinate (x1) at (200:2);
    \coordinate (x2) at (160:2);
    \coordinate (x4) at (20:2);
    \coordinate (x5) at (-20:2);
    \coordinate (x3) at (90:2);
    \coordinate (y) at (180:1.3);
    \coordinate (yp) at (0:1.3);
	\draw[thick] (x1)--(y)--(x2);
	\draw[thick] (x4)--(yp)--(x5);
	\draw[thick] (y)--(yp);
	\draw (x1) node[anchor=north east] {${\cal O}_1$};
	\draw (x2) node[anchor=south east] {${\cal O}_2$};
	\draw (x4) node[anchor=south west] {${\cal O}_3$};
	\draw (x5) node[anchor=north west] {${\cal O}_4$};
	\draw  ($(y)!0.5!(yp)$) node[anchor=north] {$\Delta$};
	\draw (y) node[vint] {};
	\draw (yp) node[vint] {};
\end{tikzpicture}}
\newsavebox{\figFourPtB}
\savebox{\figFourPtB}{%
\begin{tikzpicture}[scale=.8]
    \tikzstyle{vint}=[draw,scale=0.55,color=teal,fill=green,circle]
    \draw[thick, color=black!30] (0,0) circle[radius=2];
    \coordinate (x1) at (200:2);
    \coordinate (x2) at (160:2);
    \coordinate (x4) at (20:2);
    \coordinate (x5) at (-20:2);
    \coordinate (x3) at (90:2);
    \coordinate (y) at (180:1.3);
    \coordinate (yp) at (0:1.3);
    \coordinate (yT) at (180:.5);
    \coordinate (ypT) at (0:.5);
	\draw[thick, color=red, dashed] (x2) to [out =  -15, in = 15, looseness = 1.5] (x1);
	\draw[thick, color=red, dashed] (x4) to [out =  195, in = 165, looseness = 1.5] (x5);
	\draw[thick] (x1)--(y)--(x2);
	\draw[thick] (x4)--(yp)--(x5);
	\draw[thick] (y)--(yp);
	\draw (x1) node[anchor=north east] {${\cal O}_1$};
	\draw (x2) node[anchor=south east] {${\cal O}_2$};
	\draw (x4) node[anchor=south west] {${\cal O}_3$};
	\draw (x5) node[anchor=north west] {${\cal O}_4$};
	\draw  ($(y)!0.5!(yp)$) node[anchor=north] {$\Delta$};
	\draw  ($(y)!0.5!(yT)$) node[anchor=north] {$\Delta_{k_1}$};
	\draw  ($(yp)!0.4!(ypT)$) node[anchor=north] {$\Delta_{k_2}$};
	\draw (yT) node[vint] {};
	\draw (ypT) node[vint] {};
\end{tikzpicture}}
\newsavebox{\figFourPtC}
\savebox{\figFourPtC}{%
\begin{tikzpicture}[scale=.8]
    \tikzstyle{vint}=[draw,scale=0.55,color=teal,fill=green,circle]
    \draw[thick, color=black!30] (0,0) circle[radius=2];
    \coordinate (x1) at (200:2);
    \coordinate (x2) at (160:2);
    \coordinate (x4) at (20:2);
    \coordinate (x5) at (-20:2);
    \coordinate (x3) at (90:2);
    \coordinate (y) at (180:1.3);
    \coordinate (yp) at (0:1.3);
	\draw[thick, color=red, dashed] (x2) to [out =  -15, in = 15, looseness = 1.5] (x1);
	\draw[thick, color=red, dashed] (x4) to [out =  195, in = 165, looseness = 1.5] (x5);
	\draw[thick] (x1)--(y)--(x2);
	\draw[thick] (x4)--(yp)--(x5);
	\draw[thick] (y)--(yp);
	\draw (x1) node[anchor=north east] {${\cal O}_1$};
	\draw (x2) node[anchor=south east] {${\cal O}_2$};
	\draw (x4) node[anchor=south west] {${\cal O}_3$};
	\draw (x5) node[anchor=north west] {${\cal O}_4$};
	\draw  ($(y)!0.5!(yp)$) node[anchor=north] {$\Delta$};
\end{tikzpicture}}
\newsavebox{\figFivePtA}
\savebox{\figFivePtA}{%
\begin{tikzpicture}[scale=.8]
    \tikzstyle{vint}=[draw,scale=0.55,color=teal,fill=green,circle]
    \draw[thick, color=black!30] (0,0) circle[radius=2];
    \coordinate (x1) at (200:2);
    \coordinate (x2) at (160:2);
    \coordinate (x4) at (20:2);
    \coordinate (x5) at (-20:2);
    \coordinate (x3) at (90:2);
    \coordinate (y) at (180:1.3);
    \coordinate (yp) at (0:1.3);
    \coordinate (z) at (0:0);
	\draw[thick] (x1)--(y)--(x2);
	\draw[thick] (x4)--(yp)--(x5);
	\draw[thick] (y)--(z)--(yp);
	\draw[thick] (z)--(x3);
	\draw (x1) node[anchor=north east] {${\cal O}_1$};
	\draw (x2) node[anchor=south east] {${\cal O}_2$};
	\draw (x3) node[anchor=south] {${\cal O}_3$};
	\draw (x4) node[anchor=south west] {${\cal O}_4$};
	\draw (x5) node[anchor=north west] {${\cal O}_5$};
	\draw  ($(y)!0.5!(z)$) node[anchor=north] {$\Delta$};
	\draw  ($(yp)!0.5!(z)$) node[anchor=north] {$\Delta^\prime$};
	\draw (y) node[vint] {};
	\draw (yp) node[vint] {};
	\draw (z) node[vint] {};
\end{tikzpicture}}
\newsavebox{\figFivePtB}
\savebox{\figFivePtB}{%
\begin{tikzpicture}[scale=.8]
    \tikzstyle{vint}=[draw,scale=0.55,color=teal,fill=green,circle]
    \draw[thick, color=black!30] (0,0) circle[radius=2];
    \coordinate (x1) at (200:2);
    \coordinate (x2) at (160:2);
    \coordinate (x4) at (20:2);
    \coordinate (x5) at (-20:2);
    \coordinate (x3) at (90:2);
    \coordinate (y) at (180:1.3);
    \coordinate (yp) at (0:1.3);
    \coordinate (yT) at (180:.65);
    \coordinate (ypT) at (0:.65);
    \coordinate (z) at (0:0);
	\draw[thick] (x1)--(y)--(x2);
	\draw[thick] (x4)--(yp)--(x5);
	\draw[thick] (y)--(z)--(yp);
	\draw[thick] (z)--(x3);
	\draw[thick, color=red, dashed] (x2) to [out =  -15, in = 15, looseness = 1.5] (x1);
	\draw[thick, color=red, dashed] (x4) to [out =  195, in = 165, looseness = 1.5] (x5);
	\draw (x1) node[anchor=north east] {${\cal O}_1$};
	\draw (x2) node[anchor=south east] {${\cal O}_2$};
	\draw (x3) node[anchor=south] {${\cal O}_3$};
	\draw (x4) node[anchor=south west] {${\cal O}_4$};
	\draw (x5) node[anchor=north west] {${\cal O}_5$};
	\draw  ($(yT)!0.5!(z)$) node[anchor=north] {${\scriptstyle \Delta}$};
	\draw  ($(ypT)!0.5!(z)$) node[anchor=north] {${\scriptstyle \Delta^\prime}$};
	\draw  ($(y)!0.5!(yT)$) node[anchor=north] {${\scriptstyle \Delta_{k_1}}$};
	\draw  ($(ypT)!0.6!(yp)$) node[anchor=north] {${\scriptstyle \Delta_{k_2}}$};
	\draw (yT) node[vint] {};
	\draw (ypT) node[vint] {};
	\draw (z) node[vint] {};
\end{tikzpicture}}
\newsavebox{\figFivePtC}
\savebox{\figFivePtC}{%
\begin{tikzpicture}[scale=.8]
    \tikzstyle{vint}=[draw,scale=0.55,color=teal,fill=green,circle]
    \draw[thick, color=black!30] (0,0) circle[radius=2];
    \coordinate (x1) at (200:2);
    \coordinate (x2) at (160:2);
    \coordinate (x4) at (20:2);
    \coordinate (x5) at (-20:2);
    \coordinate (x3) at (90:2);
    \coordinate (y) at (180:1.3);
    \coordinate (yp) at (0:1.3);
    \coordinate (z) at (0:0);
	\draw[thick] (x1)--(y)--(x2);
	\draw[thick] (x4)--(yp)--(x5);
	\draw[thick] (y)--(z)--(yp);
	\draw[thick] (z)--(x3);
	\draw[thick, color=red, dashed] (x2) to [out =  -15, in = 15, looseness = 1.5] (x1);
	\draw[thick, color=red, dashed] (x4) to [out =  195, in = 165, looseness = 1.5] (x5);
	\draw (x1) node[anchor=north east] {${\cal O}_1$};
	\draw (x2) node[anchor=south east] {${\cal O}_2$};
	\draw (x3) node[anchor=south] {${\cal O}_3$};
	\draw (x4) node[anchor=south west] {${\cal O}_4$};
	\draw (x5) node[anchor=north west] {${\cal O}_5$};
	\draw  ($(y)!0.5!(z)$) node[anchor=north] {$\Delta$};
	\draw  ($(yp)!0.5!(z)$) node[anchor=north] {$\Delta^\prime$};
	\draw (z) node[vint] {};
\end{tikzpicture}}

\begin{document}

\title{Holographic dual of the five-point conformal block}
\authors{Sarthak Parikh\footnote{\tt sparikh@caltech.edu}}
\institution{Caltech}{Division of Physics, Mathematics and Astronomy, California Institute of Technology,\cr\hskip0.06in Pasadena, CA 91125, USA}

 \abstract{We present the holographic object which computes the five-point global conformal block in arbitrary dimensions for external and exchanged scalar operators. This object is interpreted as a weighted sum over infinitely many five-point geodesic bulk diagrams. These five-point geodesic bulk diagrams provide a generalization of their previously studied four-point counterparts. We prove our claim by showing that the aforementioned sum over geodesic bulk diagrams is the appropriate eigenfunction of the conformal Casimir operator with the right boundary conditions. This result rests on crucial inspiration from a much simpler $p$-adic version of the problem set up on the Bruhat--Tits tree.}

\maketitle

{\hypersetup{linkcolor=black}
\tableofcontents
}


\section{Introduction}
\label{INTRO}

Conformal blocks are theory independent  building blocks of conformal field theories which are fixed entirely from conformal invariance. Equipped with the CFT data (i.e.\ the spectrum and the OPE coefficients of any theory), the knowledge of conformal blocks permits the construction of all correlators in the theory.
Given their fundamental importance in CFTs, it is important to understand and investigate the holographic duals of conformal blocks in the context of AdS/CFT.
Such a task was  recently undertaken for four-point global conformal blocks in any dimension with external scalar operators~\cite{Hijano:2015zsa} (with further generalizations in Refs.~\cite{Nishida:2016vds,Castro:2017hpx,Dyer:2017zef,Chen:2017yia,Gubser:2017tsi,Kraus:2017ezw,Tamaoka:2017jce,Nishida:2018opl,Sleight:2017fpc,Bhatta:2016hpz,Bhatta:2018gjb,Das:2018ajg}) and for Virasoro blocks in AdS$_3$/CFT$_2$~\cite{Fitzpatrick:2014vua,Fitzpatrick:2015zha,Hijano:2015rla,Alkalaev:2015wia,Hijano:2015qja,Alkalaev:2015lca,Alkalaev:2015fbw,Banerjee:2016qca,Besken:2016ooo,Alkalaev:2016rjl,Alkalaev:2018nik}.

Nevertheless, the holographic interpretation of higher-point global conformal blocks has remained unresolved so far, and fraught with various technical obstructions. 
With the hope that new tools may provide new clues, in this paper, we appeal to the framework  of $p$-adic AdS/CFT~\cite{Gubser:2016guj,Heydeman:2016ldy,Gubser:2017tsi} which offers a significantly less complicated setting for tackling the same problem.
Indeed, it turns out, the problem of various higher-point blocks is relatively easily addressed over the $p$-adics~\cite{JP:inprep}, and owing to the close ties with the usual AdS/CFT setup (see, e.g., Refs.~\cite{Gubser:2016guj,Heydeman:2016ldy,Gubser:2017tsi,Gubser:2017vgc,Gubser:2017pyx,Jepsen:2018dqp,Jepsen:2018ajn,Gubser:2016htz,Gubser:2017qed,Gubser:2018cha,Heydeman:2018qty}) we are able to extract key new insights into the analogous calculation over reals.
 The $p$-adic results generalize the result of Ref.~\cite{Gubser:2017tsi}, and are discussed in Ref.~\cite{JP:inprep}. The $p$-adic five-point case, related to the subject of this paper, is briefly summarized later in this section. 

In this paper, inspired by the $p$-adic result, we establish the holographic dual of the global five-point block with external scalar operators in the usual (real) AdS$_{n+1}$/CFT$_n$ setup, generalizing the result of Ref.~\cite{Hijano:2015zsa} to obtain a holographic object, expressed as a \emph{weighted sum of five-point geodesic bulk diagrams}, which directly computes the five-point block. 
We present such an object in section~\ref{BLOCKDUAL}, where we also provide a proof by conformal Casimir equation in support of our claim, and close with further comments and future directions in section~\ref{DISCUSSION}.
In the remainder of this section we discuss the general strategy utilized in this paper to obtain the holographic dual of the five-point block.

\vspace{2em}

The four-point function of local scalar operators in any CFT is fixed from conformal invariance up to an arbitrary function of conformal cross-ratios. Performing, for example, an OPE in the $s$-channel, we may write
\eqn{OOOO}{
\langle {\cal O}_1(x_1) \cdots {\cal O}_4(x_4) \rangle = \sum_{{\cal O}} C_{12{\cal O}} C_{34{\cal O}} W_{\Delta,\ell}^{\Delta_1,\ldots,\Delta_4}(x_i)\,,
}
where $C_{ijk}$ are the OPE coefficients and $W_{\Delta,\ell}^{\Delta_1,\ldots,\Delta_4}(x_i)$ is the conformal partial wave which represents the contribution from conformal families with highest weight representation $(\Delta,\ell)$.\footnote{For instance, for $\ell=0$, the conformal partial wave is given by
\eqn{confwave4}{
W_{\Delta_a}^{\Delta_1,\Delta_2,\Delta_3,\Delta_4}(x_i) &= W_0^{(4)}(x_i) {\cal G}_{\Delta_a}^{\Delta_1,\Delta_2,\Delta_3,\Delta_4}(u_i) \cr 
W_0^{(4)}(x_1,x_2,x_3,x_4) &= {1 \over (x_{12}^2)^{\Delta_{12,}} (x_{34}^2)^{\Delta_{34,}}}
\left({x_{24}^2 \over x_{14}^2} \right)^{\Delta_{1,2}} 
\left({x_{14}^2 \over x_{13}^2} \right)^{\Delta_{3,4}}, 
}
where ${\cal G}_{\Delta_a}^{\Delta_1,\Delta_2,\Delta_3,\Delta_4}(u_i)$ is the four-point conformal block, given in any spacetime dimension by~\cite{Dolan:2000ut}
\eqn{calG4}{ 
{\cal G}_{\Delta_a}^{\Delta_1,\Delta_2,\Delta_3,\Delta_4}(u_1,u_2) = u_1^{\Delta_a/2} \sum_{m_1,m_2=0}^\infty {u_1^{m_1}  (1-u_2)^{m_2} \over m_1!m_2!} 
{(\Delta_{a1,2})_{m_1} (\Delta_{a4,3})_{m_1} \over (\Delta_a -n/2+1)_{m_1}} {(\Delta_{a2,1})_{m_1+m_2} (\Delta_{a3,4})_{m_1+m_2} \over (\Delta_a)_{2m_1+m_2}}
}
with the conformal cross-ratios
\eqn{u1u2}{
u_1 = {x_{12}^2 x_{34}^2 \over x_{13}^2 x_{24}^2} \qquad u_2 = {x_{14}^2 x_{23}^2 \over x_{13}^2 x_{24}^2}\,.
}
Whenever we show multiple entries in the subscript of any conformal dimension, we are using the convention~\eno{DeltaijkDef} to write linear combinations of dimensions compactly.
The Pochhammer symbol appearing above and in many subsequent equations is defined as follows:
\eqn{pochhammer}{
(a)_m \equiv {\Gamma(a+m) / \Gamma(a)}\,.
}
}
In the rest of this paper, we will suppress the angular momentum subscript $\ell$.\footnote{For simplicity we set $\ell=0$, but many of the results presented below should admit a straightforward generalization to account for the exchange of representations with non-zero $\ell$.}

As noted above, recently, Ref.~\cite{Hijano:2015zsa} advanced a holographic object which computes the four-point conformal partial wave and in turn, provides an extremely efficient computational technique for holographically obtaining  the conformal block decomposition of four-point bulk (Witten) diagrams without having to do any explicit bulk integrals.

\begin{figure}[!t]
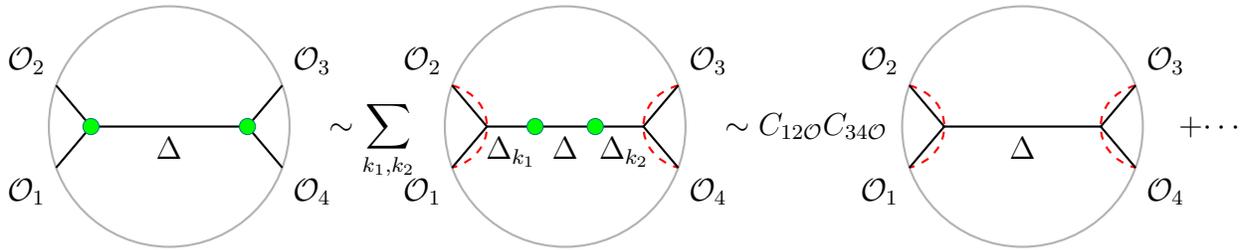

    \centering
    \[ \musepic{\figFourPtA} \!\!\!\! \sim  \sum_{k_1,k_2}\!\!\!\!\! \musepic{\figFourPtB}\!\!\!\! \sim C_{12{\cal O}} C_{34{\cal O}}\!\!\!\!\!\!\! \musepic{\figFourPtC} \!\!\! + \cdots \]
    \caption{
    Obtaining the conformal block decomposition of the four-point exchange diagram~\cite{Hijano:2015zsa}. Green, disk-shaped bulk vertices are to be integrated over all of AdS.
    The remaining bulk vertices are integrated over boundary anchored geodesics, shown as dashed, red curves. Just for this figure, we have used the shorthand: $\Delta_{k_1} = \Delta_1+\Delta_2+2k_1, \Delta_{k_2} = \Delta_3+\Delta_4+2k_2$. We used~\eno{KKexpansion} to get to the second step, and~\eno{GG} to obtain the third step. The ellipsis in the third step corresponds to contributions from double-trace exchanges.
    }
    \label{fig:4pt}
\end{figure}

Turning this around, given the CFT data, i.e.\ the spectrum and the OPE coefficients, one can hope to use the conformal block decomposition~\eno{OOOO} to extract the universal, theory independent conformal partial waves.
Heuristically (see also figure~\ref{fig:4pt}): 
\begin{itemize} 
\item For a tree-level  four-point bulk exchange diagram, one employs an AdS propagator identity~\cite{Hijano:2015zsa}  (refer to~\eno{KKexpansion}) which expresses a product of bulk-to-boundary propagators incident on a common bulk point as a series expansion in bulk solutions of the Klein-Gordon equation sourced on a boundary anchored geodesic which is, in a sense, the bulk dual of taking an OPE in the CFT~\cite{Czech:2016xec,deBoer:2016pqk,daCunha:2016crm,Guica:2016pid}.
\item To be able to extract the conformal partial wave, one still needs to evaluate the original AdS integration over all bulk interaction vertices in the diagram. This is achieved with the help of a propagator identity which expresses the integral over a product of bulk-to-bulk propagators sharing a bulk point as an un-integrated linear combination of bulk-to-bulk propagators (i.e.\ schematically, $\int_z \hat{G}(a,z)\hat{G}(b,z) \sim \hat{G}(a,b)-\hat{G}(a,b)$, see~\eno{GG})~\cite{Hijano:2015zsa}.
\item At this point, one compares with the CFT expansion~\eno{OOOO} and with prior knowledge of the OPE coefficients, one can isolate a natural candidate for the universal conformal partial wave, more precisely, the theory independent bulk object which directly computes the conformal partial wave. 
\end{itemize}
Importantly, we stress that the inverse problem of extracting  conformal blocks from~\eno{OOOO} given the OPE coefficients and the four-point function as outlined here is genuinely non-unique. It is used merely as a heuristic guide to obtain a natural candidate for the conformal block, which must then be verified using an independent check. Luckily, it turns out for the examples considered in this paper, this strategy leads to the best case scenario and does indeed provide the desired holographic objects which compute the conformal partial waves,  as confirmed by a proof by Casimir equation. 

Essentially, all we have described above is simply a holographic strategy to obtain the projection of the four-point function, to the conformal family of the operator ${\cal O}$ of dimension $\Delta$:
\eqn{ProjectConf4}{
W_\Delta^{\Delta_1,\ldots,\Delta_4}(x_i) = {1 \over C_{12{\cal O}} C_{34{\cal O}}} \langle  {\cal O}_1(x_1) {\cal O}_2(x_2) P_\Delta {\cal O}_3(x_3) {\cal O}_4(x_4) \rangle \,,
}
where the projection operator is given by
\eqn{Projector}{
P_\Delta = \sum_k | P^k {\cal O} \rangle \langle P^k {\cal O}|\,,
}
where $P^k$ stands for $k$ applications of the momentum generators $P_\mu$ to obtain the descendants of ${\cal O}$.

The upshot of this exercise is that the following  geodesic bulk integral, 
\eqn{4ptGeodesic}{
{\cal W}_{\Delta}^{\Delta_1,\ldots,\Delta_4}(x_i) \equiv  \displaystyle{\iint_{\substack{w \in \gamma_{12}\\ w^\prime \in \gamma_{34}}} 
\hat{K}_{\Delta_1}(x_1,w) \hat{K}_{\Delta_2}(x_2,w) 
\hat{G}_{\Delta}(w,w^\prime)
\hat{K}_{\Delta_3}(x_3,w^\prime) \hat{K}_{\Delta_4}(x_4,w^\prime)}\,,
} 
where $\hat{G}, \hat{K}$ are bulk-to-bulk and bulk-to-boundary propagators respectively,
computes the conformal partial wave associated with the exchange of an operator of dimension $\Delta$ between external insertions ${\cal O}_1,{\cal O}_2$ and ${\cal O}_3, {\cal O}_4$~\cite{Hijano:2015zsa} (see figure~\ref{fig:calW4}).
\begin{figure}[!t]
    \centering
    \[ {\cal W}_{\Delta}^{\Delta_1,\Delta_2,\Delta_3,\Delta_4}(x_1,x_2,x_3,x_4) = \musepic{\figcalWfour} \]
    \caption{The geodesic integral of~\eno{4ptGeodesic}, represented diagrammatically as a geodesic bulk diagram.
    The dashed, red curves represent the boundary anchored geodesics along which the respective bulk interaction vertices are integrated.
    In the literature, this is often referred to as a (four-point) geodesic Witten diagram~\cite{Hijano:2015zsa}. In particular, the same diagram computes the $p$-adic four-point conformal partial wave when evaluated on the Bruhat--Tits tree instead of AdS~\cite{Gubser:2017tsi}. (In later figures, locations of operator insertions will often be suppressed when no confusion is likely.)}
    \label{fig:calW4}
\end{figure}
Here we are using the notation,
\eqn{}{
\int_{w \in \gamma_{ij}} \equiv \int_{-\infty}^\infty d\lambda\,,
}
where the bulk point $w=w(\lambda)$ is parametrized by  $\lambda$ running along the boundary anchored geodesic $\gamma_{ij}$ between boundary points $x_i$ and $x_j$.
More precisely, it can be shown that~\cite{Hijano:2015zsa}
\eqn{calWconfwave4}{
{\cal W}_{\Delta_a}^{\Delta_1,\Delta_2,\Delta_3,\Delta_4}(x_i) = {1\over 4}\:
\beta_\infty(2\Delta_{a1,2},2\Delta_{a2,1})\: \beta_\infty(2\Delta_{a3,4},2\Delta_{a4,3})\:
W_{\Delta_a}^{\Delta_1,\Delta_2,\Delta_3,\Delta_4}(x_i)\,,
}
where $W_{\Delta_a}^{\Delta_1,\ldots,\Delta_4}$ takes the explicit form given in~\eno{confwave4}-\eno{u1u2}, $\beta_\infty(\cdot,\cdot)$ is related to the usual Euler Beta function and defined as
\eqn{betainftyDef}{
\beta_\infty(s,t) \equiv {\zeta_\infty(s)\zeta_\infty(t) \over \zeta_\infty(s+t) } = {\Gamma(s/2) \Gamma(t/2) \over \Gamma((s+t)/2)}\,,
}
where the local zeta function $\zeta_\infty$ is defined via
\eqn{zetainftyDef}{
\zeta_\infty(s) = \pi^{-s/2}\: \Gamma(s/2) \,,
}
and we are employing the convention
\eqn{DeltaijkDef}{
\Delta_{i_1\ldots i_\ell,i_{\ell+1} \ldots i_k} \equiv {1\over 2} \left( \Delta_{i_1} + \cdots + \Delta_{i_\ell} - \Delta_{i_{\ell+1}} - \cdots - \Delta_{i_k} \right).
}

\begin{figure}[!t]
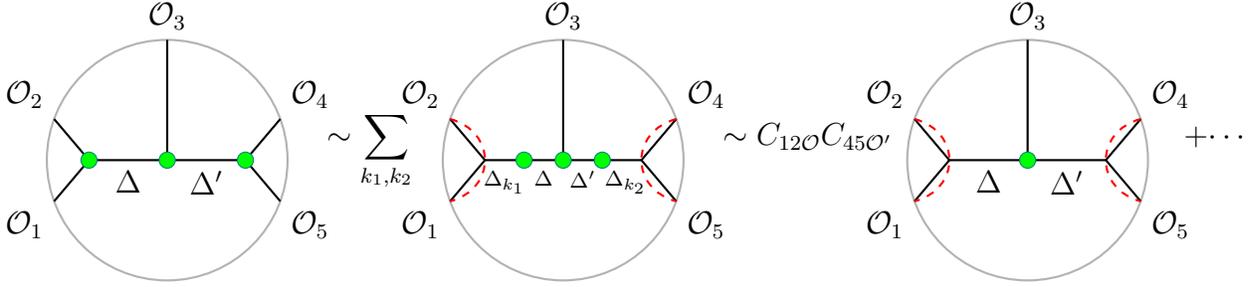

    \centering
    \[ \musepic{\figFivePtA} \!\!\!\! \sim  \sum_{k_1,k_2}\!\!\!\!\! \musepic{\figFivePtB}\!\!\!\! \sim C_{12{\cal O}} C_{45{\cal O^\prime}}\!\!\!\!\!\!\! \musepic{\figFivePtC} \!\!\! + \cdots \]
    \caption{
    Attempting a conformal block decomposition of a five-point bulk diagram. Just for this figure, we have used the shorthand: $\Delta_{k_1} = \Delta_1+\Delta_2+2k_1, \Delta_{k_2} = \Delta_4+\Delta_5+2k_2$. The figure in the third step shows the unevaluated $\int \hat{K}\hat{G}\hat{G}$ AdS integral.
    }
    \label{fig:5pt}
\end{figure}

The procedure sketched above applies more generally to higher-point blocks as well. In this paper, we focus on the five-point case. To obtain the holographic dual of the five-point conformal partial wave defined via the projection,
\eqn{ProjectConf5}{
W_{\Delta,\Delta^\prime}^{\Delta_1,\ldots,\Delta_5}(x_i) = {1 \over C_{12{\cal O}} C_{{\cal O} 3 {\cal O}^\prime} C_{45{\cal O}^\prime}} \langle  {\cal O}_1(x_1) {\cal O}_2(x_2) P_\Delta {\cal O}_3(x_3) P_{\Delta^\prime} {\cal O}_4(x_4) {\cal O}_5(x_5) \rangle\,,
}
we proceed as follows (see also figure~\ref{fig:5pt}):
\begin{itemize}
    \item We start with a five-point bulk diagram with two single-trace exchanges in the intermediate channels, and once again apply the holographic OPE identity~\eno{KKexpansion} to operator insertions ${\cal O}_1(x_1) {\cal O}_2(x_2)$ and ${\cal O}_4(x_4) {\cal O}_5(x_5)$. This is expected to account for two  of the OPE coefficients in the denominator of~\eno{ProjectConf5}. We will return to the final OPE coefficient $C_{{\cal O}3{\cal O}^\prime}$ shortly.
    \item There are still three bulk interaction vertices to integrate over all of AdS, out of which two integrations can be carried out using the previously mentioned $\int \hat{G}\hat{G} \sim \hat{G}-\hat{G}$ identity~\eno{GG}. This leaves a final integral over a combination of two bulk-to-bulk propagators and one bulk-to-boundary propagator sharing a common vertex, to be integrated over all of AdS, i.e.\ an integral of the form $\int_z \hat{K}_{\Delta_3}(x_3,z)\hat{G}_\Delta(a,z)\hat{G}_{\Delta^\prime}(b,z)$.
    \item  In particular, we are interested in the contribution to the $\int \hat{K}\hat{G}\hat{G}$ integral which is proportional to the final OPE coefficient $C_{{\cal O}3{\cal O}^\prime}$, since this contribution  isolates precisely the holographic term which computes the theory independent
    partial wave~\eno{ProjectConf5}.
\end{itemize}

However, this procedure hits a roadblock in the third step, since an AdS propagator identity of the form $\int_z \hat{K}_{\Delta_3}(x_3,z)\hat{G}_\Delta(a,z)\hat{G}_{\Delta^\prime}(b,z)$ is not readily available.
At this point, we appeal to the $p$-adic formulation, where the analogous $\int \hat{K} \hat{G} \hat{G}$ identity is easily obtained and was, in fact, already written down in Ref.~\cite{Gubser:2017tsi}. We reproduce it below in a  mathematically equivalent, but highly suggestive form~\cite{JP:inprep}:\footnote{The identity~\eno{GGK} has been written in terms of an OPE coefficient, bulk scalar mass, and a normalization factor since it turns out that real propagator identities may also be written in terms of the same quantities. The details of the exact form these coefficients take is not important for our purposes so we only make a brief comment: The $p$-adic analog of the mass-dimension relation~\cite{Gubser:2016guj} is $m_\Delta^2 = -1/(\zeta_p(-\Delta)\zeta_p(\Delta-n))$ with the local zeta function $\zeta_p$ defined in~\eno{betazetapDef}. The coefficient $N_{\Delta} =  { \zeta_p(2\Delta) / p^{\Delta}}$ is related to the choice of normalization of the bulk-to-bulk propagator via its equation of motion on the Bruhat--Tits tree, similar to~\eno{GhatEOM} below for the AdS bulk-to-bulk propagator (see, e.g.\ Ref.~\cite{Jepsen:2018ajn}).}
\eqn{GGK}{ 
\sum_{z\in {\cal T}_{p^n}} \hat{K}_{\Delta_3}(x_3,z) \hat{G}_{\Delta_a}(a,z) \hat{G}_{\Delta_b}(b,z)  &= C_{3ab}\: \hat{K}_{\Delta_{3a,b}}(x_3,a)\hat{K}_{\Delta_{3b,a}}(x_3,b)\hat{G}_{\Delta_{ab,3}}(a,b) \cr 
 &- {1 \over m^2_{2\Delta_{a3,}}-m^2_{\Delta_b}}{1 \over N_{\Delta_b}} \hat{G}_{\Delta_a}(a,b) \hat{K}_{\Delta_3}(x_3,b) \cr 
 &- {1 \over m^2_{2\Delta_{b3,}}-m^2_{\Delta_a}}{1\over N_{\Delta_a}} \hat{G}_{\Delta_b}(a,b) \hat{K}_{\Delta_3}(x_3,a) \,,
}
where the discrete sum is over all vertices of the $(p^n+1)$-regular Bruhat--Tits tree ${\cal T}_{p^n}$ (the discrete, $p$-adic analog of AdS space). The coefficient $C_{ijk}$ is the OPE coefficient of the dual generalized free-field theory, whose explicit form is unimportant. This is because we are  after only the theory independent conformal partial wave whose holographic representation is dictated by the particular combination of bulk-to-bulk and bulk-to-boundary propagators as they appear in the term proportional to the OPE coefficient $C_{3ab}$ above.

\begin{figure}[!t]
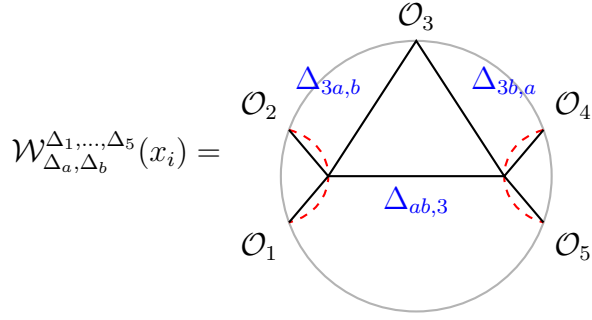

    \centering
    \[ {\cal W}_{\Delta_a,\Delta_b}^{\Delta_1,\ldots,\Delta_5}(x_i) = \musepic{\figcalWpadic} \]
    \caption{The five-point geodesic bulk diagram \emph{on the Bruhat--Tits tree} dual to the five-point $p$-adic conformal partial wave~\cite{JP:inprep}.}
    \label{fig:padicBlock}
\end{figure}

This motivates the five-point proposal~\cite{JP:inprep} depicted in figure~\ref{fig:padicBlock}, which is a direct generalization of the four-point geodesic bulk diagram in figure~\ref{fig:calW4}.
Indeed, a direct computation on the Bruhat--Tits tree confirms the five-point geodesic bulk diagram (figure~\ref{fig:padicBlock}) computes the ($p$-adic) five-point conformal block up to an overall factor~\cite{JP:inprep},
\eqn{calWpadic}{
{\cal W}_{\Delta_a,\Delta_b}^{\Delta_1,\ldots,\Delta_5}(x_i) &=   \beta_p(2\Delta_{a1,2},2\Delta_{a2,1})\: \beta_p(2\Delta_{b4,5},2\Delta_{b5,4}) \:
W_{\Delta_a,\Delta_b}^{\Delta_1,\ldots,\Delta_5}(x_i) \cr 
 &= \beta_p(2\Delta_{a1,2},2\Delta_{a2,1})\: \beta_p(2\Delta_{b4,5},2\Delta_{b5,4}) \: W_0^{(5)}(x_i)\: u^{\Delta_a/2} v^{\Delta_b/2}
}
where
\eqn{betazetapDef}{
\beta_p(s,t) \equiv {\zeta_p(s)\zeta_p(t) \over \zeta_p(s+t) } \qquad \qquad  \zeta_p(s) = {1 \over 1-p^{-s}}\,,
}
and $W_0^{(5)}(x_i)$ is the $p$-adic analog of the five-point ``leg factor'' defined below in~\eno{W05Def} for the real case.
Since the dual  (generalized free-field) $p$-adic CFT has only primary operators in its spectrum with no descendants~\cite{Melzer:1988he,Gubser:2017tsi}, all conformal blocks are trivial, as instantiated in the second line of~\eno{calWpadic} for the five-point block, where $u, v$ are the two independent conformal cross-ratios\footnote{There are only two independent cross-ratios since the boundary of the Bruhat--Tits tree, given by the projective line over the $p$-adic numbers, is one-dimensional. Recall that in a one-dimensional (real) CFT, there is only one independent cross-ratio for the four-point block, and exactly two for the five-point block.} 
\eqn{uvDef}{
u \equiv \left|{ x_{12}^2 x_{34}^2 \over x_{13}^2x_{24}^2}\right|_p \qquad v \equiv \left| {x_{23}^2 x_{45}^2 \over x_{24}^2 x_{35}^2 }\right|_p,
}
with $|\cdot|_p$ denoting the $p$-adic norm.

The $p$-adic propagator identity~\eno{GGK} and the geodesic bulk diagram in figure~\ref{fig:padicBlock} anticipate the corresponding candidates for the real analog. 
Indeed, the contribution to the (real) $\int \hat{K}\hat{G}\hat{G}$ integral, proportional to the structure constant of the generalized free-field theory, takes the form~\cite{JP:inprep},
\eqn{GGKrealfijk}{
& \int_{z \in {\rm AdS}} \hat{K}_{\Delta_3}(x_3,z)\hat{G}_{\Delta_a}(a,z)\hat{G}_{\Delta_b}(b,z) \cr 
&\sim C_{3ab} \sum_{k_1,k_2,k_3=0}^\infty c_{k_1,k_2,k_3} 
 \hat{K}_{\Delta_{3a,b}-k_1+k_2}(x_3,a)  \hat{K}_{\Delta_{3b,a}+k_1-k_2}(x_3,b)
 \hat{G}_{\Delta_{ab,3}+k_1+k_2+2k_3}(a,b) \cr 
 &\quad + \cdots\,,
}
where the ellipsis represents the remaining terms which are analogs of terms in the second and third lines of~\eno{GGK}, and the coefficient $c_{k_1,k_2,k_3}$ will be given later in~\eno{c3Def}.
As is often the case, the real analog of the $p$-adic AdS/CFT result uplifts the simpler $p$-adic result to include more complicated sums. These sums essentially account for the full descendant contribution to each conformal family, which are trivially absent over the $p$-adics.
With~\eno{GGKrealfijk} in hand, we are able to easily write down the real analog of the holographic dual shown in figure~\ref{fig:padicBlock}, which we describe in the next section.

\section{Holographic dual of the five-point conformal block}
\label{BLOCKDUAL}

The five-point conformal partial wave is proportional to the five-point conformal block ${\cal G}_{\Delta_a,\Delta_b}^{\Delta_1,\ldots,\Delta_5}(u_i)$,  up to an overall ``leg factor'' $W_0^{(5)}$ given by
\eqn{W05Def}{
W_0^{(5)}(x_i) &\equiv 
{1 \over (x_{12}^2)^{\Delta_{12,}} (x_{34}^2)^{\Delta_3/2} (x_{45}^2)^{\Delta_{45,}}}
\left({x_{23}^2 \over x_{13}^2} \right)^{\Delta_{1,2}} 
\left({x_{24}^2 \over x_{23}^2} \right)^{\Delta_{3}/2}
\left({x_{35}^2 \over x_{34}^2} \right)^{\Delta_{4,5}}, 
}
such that
\eqn{confwaveblock}{
W_{\Delta_a,\Delta_b}^{\Delta_1,\ldots,\Delta_5}(x_i) = W_0^{(5)}(x_i)\: {\cal G}_{\Delta_a,\Delta_b}^{\Delta_1,\ldots,\Delta_5}(u_i)\,,
}
with the block simply a function of conformal cross-ratios.

The five-point global block in any spacetime dimension was recently explicitly computed in Ref.~\cite{Rosenhaus:2018zqn} (see also Ref.~\cite{Alkalaev:2015fbw}).
We claim the following ``geodesic bulk diagram'',
\eqn{calWalt}{
{\cal W}_{\Delta_a,\Delta_b}^{\Delta_1,\ldots,\Delta_5}(x_i) &\equiv \sum_{k_1,k_2,k_3=0}^\infty c_{k_1,k_2,k_3} \musepic{\figcalWalt}\,,
}
with the coefficients, the same as those which appear in~\eno{GGKrealfijk},  given by
\eqn{c3Def}{
c_{k_1,k_2,k_3} &\equiv  {(-1)^{k_1+k_2+k_3} \over k_1!k_2!k_3!} 
{\left(1-\Delta_{3a,b}\right)_{k_1} \left(1-\Delta_{3b,a}\right)_{k_2} \over
\left(\Delta_b-n/2+1\right)_{k_1}
\left(\Delta_a-n/2+1\right)_{k_2}
\left(\Delta_{ab,3}-n/2+k_1+k_2+k_3\right)_{k_3}
} 
 \cr 
& \times\left( \Delta_{3a,b} \right)_{-k_1+k_2} 
\left(\Delta_{ab,3}\right)_{k_1+k_2+2k_3} 
\left(\Delta_{3b,a}\right)_{k_1-k_2} \:
\pFq{3}{2}{-k_1,-k_2,\Delta_{ab3,\!\!}-n/2}{\Delta_{3a,\!\!b}-k_1,\Delta_{3b,\!\!a}-k_2}{1} \,,
 }
 and the geodesic integral defined as,
 \eqn{BasicDiagG}{
\musepic{\figcalWGdef} \equiv \begin{matrix} \displaystyle{\iint_{\substack{w \in \gamma_{12}\\ w^\prime \in \gamma_{45}}} 
\hat{K}_{\Delta_1}(x_1,w) \hat{K}_{\Delta_2}(x_2,w) 
\hat{K}_{\Delta_4}(x_4,w^\prime) \hat{K}_{\Delta_5}(x_5,w^\prime)} \cr 
\times \hat{K}_{\Delta_L}(x_3,w) \hat{G}_{\Delta_C}(w,w^\prime)  \hat{K}_{\Delta_R}(x_3,w^\prime)
    \end{matrix}\,,
} 
 computes the five-point conformal partial wave (and hence the conformal block) up to an overall constant factor, i.e.\
\eqn{calWconfwave}{
{\cal W}_{\Delta_a,\Delta_b}^{\Delta_1,\ldots,\Delta_5}(x_i) = {1\over 4}\: 
\beta_\infty(2\Delta_{a1,2},2\Delta_{a2,1})\: \beta_\infty(2\Delta_{b4,5},2\Delta_{b5,4}) \:
W_{\Delta_a,\Delta_b}^{\Delta_1,\ldots,\Delta_5}(x_i)\,,
}
where $\beta_\infty(\cdot,\cdot)$ is defined in~\eno{betainftyDef}.
Note that setting $k_1=k_2=k_3=0$ in~\eno{calWalt} recovers the $p$-adic form of the geodesic bulk diagram given in figure~\ref{fig:padicBlock}.

 The standard scalar bulk-to-bulk propagator utilized above takes the form\footnote{In this paper we have chosen to normalize the bulk-to-bulk propagator according to
\eqn{GhatEOM}{
\left(-\nabla^2_{\rm AdS} + m_\Delta^2\right) \hat{G}_\Delta(z,z^\prime) &= -{1\over N_{\Delta}} \delta^{n+1}(z-z^\prime) \qquad N_\Delta = -{\zeta_\infty(2\Delta) \over 2 \nu_\Delta \zeta_\infty(2\Delta-n) }\,,
}
where $2\nu_\Delta = 2\Delta-n$, $\zeta_\infty$ is defined in~\eno{zetainftyDef} and $z, z^\prime$ are $(n+1)$-dimensional bulk coordinates.
The overall constant on the right hand side is merely a repackaging of the familiar normalization factor of the scalar bulk-to-bulk propagator found in the literature, in terms of the local zeta function~\eno{zetainftyDef}. We have included it explicitly in~\eno{GhatEOM} for the convenience of working with an ``unnormalized'' bulk-to-bulk propagator as written in~\eno{Gxi}.}
  \eqn{Gxi}{
 \hat{G}_{\Delta}(w,w^\prime) &= \left({\xi(w,w^\prime) \over 2} \right)^{\Delta} \pFq{2}{1}{{\Delta \over 2},{\Delta+1 \over 2}}{\Delta-n/2+1}{\xi(w,w^\prime)^2}\cr 
 &= \sum_{k=0}^\infty {1 \over k!} {(\Delta)_{2k} \over (\Delta-n/2+1)_k} \left({\xi(w,w^\prime) \over 2}\right)^{\Delta+2k}\,,
 }
where $\xi(w,w^\prime)$ is related  to the geodesic distance $\sigma$ between bulk points $w, w^\prime$ via 
 \eqn{sigmaxi}{ 
 \xi(w,w^\prime)^{-1} = \cosh \sigma(w,w^\prime) \,.
 }
 The bulk-to-boundary propagator is given by
 \eqn{Kdef}{
 \hat{K}_\Delta(x,z) = { z_0^\Delta \over (z_0^2 + |\vec{z}-\vec{x}|^2)^\Delta}\,,
 }
 where $z=(z_0,\vec{z})$ is a bulk Poincar\'{e} coordinate in (Euclidean) AdS$_{n+1}$.
The relation~\eno{Gxi} can be inverted to give,\footnote{The inverse relation~\eno{G2xi} is easy to  check analytically for $n=2$. For other values of $n$, we provide an indirect proof in appendix~\ref{PROP}.} 
\eqn{G2xi}{
\left({\xi(z,z^\prime) \over 2}\right)^\Delta = \sum_{k=0}^\infty {(-1)^k \over k!} {(\Delta)_{2k} \over (\Delta-n/2+k)_k} \hat{G}_{\Delta+2k}(z,z^\prime)\,.
}
Using the inverse relation, one can perform the $k_3$ sum in~\eno{calWalt} to get 
\eqn{calW}{
{\cal W}_{\Delta_a,\Delta_b}^{\Delta_1,\ldots,\Delta_5}(x_i) &= \sum_{k_1,k_2=0}^\infty c_{k_1,k_2} \musepic{\figcalW}\,,
}
where the bulk diagram featuring above is defined using the following geodesic integral,
\eqn{BasicDiag}{
\musepic{\figcalWdef} \equiv \begin{matrix} \displaystyle{\iint_{\substack{w \in \gamma_{12}\\ w^\prime \in \gamma_{45}}} 
\hat{K}_{\Delta_1}(x_1,w) \hat{K}_{\Delta_2}(x_2,w) 
\hat{K}_{\Delta_4}(x_4,w^\prime) \hat{K}_{\Delta_5}(x_5,w^\prime)} \cr 
\times \hat{K}_{\Delta_L}(x_3,w) \left(\displaystyle{\xi(w,w^\prime) \over 2}\right)^{\Delta_C}  \hat{K}_{\Delta_R}(x_3,w^\prime)
    \end{matrix}\,,
} 
 with the coefficients now taking the form, 
\eqn{cDef}{
c_{k_1,k_2} &\equiv  {(-1)^{k_1+k_2} \over k_1!k_2!} 
{\left(1-\Delta_{3a,b}\right)_{k_1} \left(1-\Delta_{3b,a}\right)_{k_2} \over
\left(\Delta_b-n/2+1\right)_{k_1}
\left(\Delta_a-n/2+1\right)_{k_2}} 
\left( \Delta_{3a,b} \right)_{-k_1+k_2} 
\left(\Delta_{ab,3}\right)_{k_1+k_2} 
\left(\Delta_{3b,a}\right)_{k_1-k_2} \cr 
&\qquad\qquad\qquad\qquad\qquad\qquad\qquad\qquad\qquad\qquad \times \pFq{3}{2}{-k_1,-k_2,\Delta_{ab3,\!\!}-n/2}{\Delta_{3a,\!\!b}-k_1,\Delta_{3b,\!\!a}-k_2}{1} \,.
 }
 As emphasized using the black dotted lines joining bulk points in~\eno{calW} and~\eno{BasicDiag}, the bulk-to-bulk propagator gets replaced by the simpler power-law factor $\xi(w,w^\prime)^\Delta$.

 We make four remarks:
 \begin{itemize}
     \item  Technically~\eno{calWalt} represents \emph{a sum over} geodesic bulk diagrams; by abuse of terminology we often refer to it (and thus~\eno{calW}) simply as a geodesic bulk diagram.
     \item Second, for calculational convenience, from now on  we will mostly be using the representation given in~\eno{calW}.
     \item Third, it is easily checked that~\eno{calW} reduces to the four-point geodesic bulk diagram~\cite{Hijano:2015zsa} in the limit $\Delta_3 \to 0, \Delta_b \to \Delta_a$, whereupon the double sum in~\eno{calW} collapses to a single sum over the variable $k_1=k_2$ since the coefficients $c_{k_1,k_2}$ vanish otherwise. The hypergeometric function reduces to $(\Delta_a-n/2+1)_{k_1}/k_1!$, and \eno{calW} simplifies to
\eqn{4ptlimInit}{
{\cal W}_{\Delta_a,\Delta_b}^{\Delta_1,\ldots,\Delta_5}(x_i) \stackrel{\substack{\Delta_3\to 0\\ \Delta_b\to \Delta_a}}{=}
\sum_{k_1=0}^\infty {1\over k_1!} {(\Delta_a)_{2k_1} \over (\Delta_a-n/2+1)_{k_1}} \musepic{\figcalWfourxi}\,.
}
Recognizing the remaining sum with the one shown in~\eno{Gxi}, we are led directly to the four-point geodesic bulk diagram of Ref.~\cite{Hijano:2015zsa} defined above in~\eno{4ptGeodesic},
\eqn{4ptlim}{
{\cal W}_{\Delta_a,\Delta_b}^{\Delta_1,\Delta_2,\Delta_3,\Delta_4,\Delta_5}(x_1,x_2,x_3,x_4,x_5) \stackrel{\substack{\Delta_3\to 0\\ \Delta_b\to \Delta_a}}{=} {\cal W}_{\Delta_a}^{\Delta_1,\Delta_2,\Delta_4,\Delta_5}(x_1,x_2,x_4,x_5)\,.
}

    \item Finally, the proposed relation~\eno{calWconfwave} between the five-point geodesic diagram and conformal partial wave is recognized as a simple generalization of the four-point relation~\eno{calWconfwave4}, thus providing further  evidence in support of the claim that \eno{calWalt} is an appropriate five-point generalization of the four-point geodesic bulk diagram.
 \end{itemize}

 We will now prove~\eno{calWconfwave} in the next two subsections. In section~\ref{CASIMIR} we show that the five-point geodesic bulk diagram is an eigenfunction of the conformal Casimir with the correct eigenvalues. Before proceeding with that, we first show that the conformal partial wave  as defined via the geodesic bulk diagram in~\eno{calWconfwave} has the correct boundary conditions, by investigating the OPE limit. 
 Together, these checks constitute a complete proof of~\eno{calWconfwave}~\cite{SimmonsDuffin:2012uy}.

\subsection{OPE limit}
\label{OPE}
In the limit when the boundary point $x_5 \to x_4$, we expect the five-point conformal partial wave  to reduce, to leading order, to the four-point conformal partial wave, upto an overall scaling factor as follows (see, e.g., Ref.~\cite{Rosenhaus:2018zqn})
\eqn{OPElimGen}{
W_{\Delta_a,\Delta_b}^{\Delta_1,\Delta_2,\Delta_3,\Delta_4,\Delta_5}(x_1,x_2,x_3,x_4,x_5) \stackrel{x_5 \to x_4}{\longrightarrow} (x_{45}^2)^{\Delta_{b,45}}\: W_{\Delta_a}^{\Delta_1,\Delta_2,\Delta_3,\Delta_b}(x_1,x_2,x_3,x_4)\,.
}
We will now reproduce this limit starting from the geodesic bulk diagram in~\eno{calW}. Using~\eno{calWconfwave} to obtain the conformal partial wave, sending $x_5 \to x_4$ and simultaneously evaluating the limiting geodesic integral over $y^\prime \in \gamma_{45}$ in~\eno{BasicDiag}, we obtain a double-sum over  a summand which is proportional to $(x_{45}^2)^{\Delta_{b,45}+k_1}$. The leading contribution comes from collapsing the sum over $k_1$ to the leading term at $k_1=0$, which corresponds to the contribution coming solely from a primary operator. Simplifying a bit, we obtain
\eqn{}{
&W_{\Delta_a,\Delta_b}^{\Delta_1,\Delta_2,\Delta_3,\Delta_4,\Delta_5}(x_1,x_2,x_3,x_4,x_5)  = {4\:  {\cal W}_{\Delta_a,\Delta_b}^{\Delta_1,\Delta_2,\Delta_3,\Delta_4,\Delta_5}(x_1,x_2,x_3,x_4,x_5) \over \beta_\infty(2\Delta_{a1,2},2\Delta_{a2,1})\: \beta_\infty(2\Delta_{b4,5},2\Delta_{b5,4})}
\cr 
\hspace{-0.5cm} \stackrel{x_5 \to x_4}{\longrightarrow}& {2\: (x_{45}^2)^{\Delta_{b,45}} \over \beta_\infty(2\Delta_{a1,2},2\Delta_{a2,1})}  \sum_{k_2=0}^\infty {(\Delta_{a3,b})_{k_2} (\Delta_{ab,3})_{k_2} \over k_2! (\Delta_a-n/2+1)_{k_2}}
{1 \over (x_{34}^2)^{\Delta_{b3,a}-k_2}} \musepic{\figOPElim} \,.
} 
Using the propagator identity~\eno{KKGid}, we can trade the sum over $k_2$ for a geodesic integral, to arrive at
\eqn{}{
&W_{\Delta_a,\Delta_b}^{\Delta_1,\Delta_2,\Delta_3,\Delta_4,\Delta_5}(x_1,x_2,x_3,x_4,x_5) \cr 
\stackrel{x_5 \to x_4}{\longrightarrow}&  {4\: (x_{45}^2)^{\Delta_{b,45}} \over \beta_\infty(2\Delta_{a1,2},2\Delta_{a2,1}) \beta_\infty(2\Delta_{a3,b},2\Delta_{ab,3})}  \musepic{\figcalWfourOPE} \cr 
&= (x_{45}^2)^{\Delta_{b,45}} {4\:  {\cal W}_{\Delta_a}^{\Delta_1,\Delta_2,\Delta_3,\Delta_b}(x_1,x_2,x_3,x_4) \over \beta_\infty(2\Delta_{a1,2},2\Delta_{a2,1}) \beta_\infty(2\Delta_{a3,b},2\Delta_{ab,3})} \cr 
&= (x_{45}^2)^{\Delta_{b,45}} \: W_{\Delta_a}^{\Delta_1,\Delta_2,\Delta_3,\Delta_b}(x_1,x_2,x_3,x_4)\,,
}
where we used~\eno{calWconfwave4} in the final step. This establishes the OPE limit~\eno{OPElimGen}. An identical analysis in the limit $x_2 \to x_1$ leads once again to the expected behaviour. Thus we have demonstrated that the geodesic bulk diagram~\eno{calW} (and the conformal partial wave as defined by~\eno{calWconfwave}) has the right boundary conditions. Next we show that the five-point geodesic bulk diagram is an eigenfunction of the conformal Casimir with the correct eigenvalues, which completes the proof of the claim that \eno{calW} is proportional to the five-point conformal partial wave precisely as proposed in~\eno{calWconfwave}.

\subsection{Proof by conformal Casimir equation} 
\label{CASIMIR}

Let ${\cal L}_{AB}$ denote the generators of $SO(n+1,1)$, the $n$-dimensional conformal group of a Euclidean CFT. Then, the Casimir of the conformal group is the quadratic combination ${\cal L}^2 \equiv {1\over 2} {\cal L}_{AB} {\cal L}^{AB}$. 
Conformal partial waves are eigenfunctions of the quadratic Casimir acting on multiple points. For instance, for the four-point scalar conformal partial wave~\cite{Dolan:2003hv},
\eqn{Casimir4}{
\left({\cal L}^{(1)} + {\cal L}^{(2)}\right)^2
W_{\Delta_a}^{\Delta_1,\ldots,\Delta_4}(x_1,x_2,x_3,x_4) = C_2(\Delta_a) W_{\Delta_a}^{\Delta_1,\ldots,\Delta_4}(x_1,x_2,x_3,x_4)\,,
}
where we have defined
\eqn{Lijk}{
\left( {\cal L}^{(i_1)} + \cdots + {\cal L}^{(i_k)} \right)^2 \equiv {1\over 2} \left({\cal L}^{(i_1)}_{AB} +\cdots+ {\cal L}^{(i_k)}_{AB} \right) \left({\cal L}^{(i_1)}{}^{AB} +\cdots+ {\cal L}^{(i_k)}{}^{AB} \right),
}
where ${\cal L}^{(i)}_{AB}$ is a differential operator acting on and built from the position coordinate $x_i$,
and\footnote{As previously noted,  for simplicity we are focusing only on scalar exchanges. The equations presented here generalize easily to the case of symmetric-traceless exchanges of spin $\ell$.}
\eqn{CDelta}{
C_2(\Delta) = m^2_\Delta= \Delta(\Delta-n)\,.
}
For the five-point scalar conformal partial wave~\eno{ProjectConf5}~\cite{Alkalaev:2015fbw,Rosenhaus:2018zqn}
\eqn{Casimir5}{
\left({\cal L}^{(1)} + {\cal L}^{(2)}\right)^2
W_{\Delta_a,\Delta_b}^{\Delta_1,\ldots,\Delta_5}(x_1,x_2,x_3,x_4,x_5) &= C_2(\Delta_a)\: W_{\Delta_a,\Delta_b}^{\Delta_1,\ldots,\Delta_5}(x_1,x_2,x_3,x_4,x_5) \cr 
\left({\cal L}^{(4)} + {\cal L}^{(5)}\right)^2
W_{\Delta_a,\Delta_b}^{\Delta_1,\ldots,\Delta_5}(x_1,x_2,x_3,x_4,x_5) &= C_2(\Delta_b)\: W_{\Delta_a,\Delta_b}^{\Delta_1,\ldots,\Delta_5}(x_1,x_2,x_3,x_4,x_5)\,.
}
We note that for the five-point block, the action of $\left({\cal L}^{(1)} + {\cal L}^{(2)}\right)^2$ is the same as the action of $\left({\cal L}^{(3)} + {\cal L}^{(4)} + {\cal L}^{(5)} \right)^2$, and similarly for $\left({\cal L}^{(4)} + {\cal L}^{(5)}\right)^2$.
We will now explicitly demonstrate that the geodesic bulk diagram in~\eno{calW} satisfies the second equation above, i.e.\
\eqn{calWCasimir}{
\left({\cal L}^{(1)} + {\cal L}^{(2)} + {\cal L}^{(3)}\right)^2 {\cal W}_{\Delta_a,\Delta_b}^{\Delta_1,\ldots,\Delta_5}(x_1,x_2,x_3,x_4,x_5) &= C_2(\Delta_b)\: {\cal W}_{\Delta_a,\Delta_b}^{\Delta_1,\ldots,\Delta_5}(x_1,x_2,x_3,x_4,x_5)\,.
}
The geodesic diagram~\eno{calW} is manifestly symmetric under the exchange $(1,2,3,4,5) \leftrightarrow (5,4,3,2,1)$ and $(\Delta_a,\Delta_b) \leftrightarrow (\Delta_b,\Delta_a)$. Thus the first equation of~\eno{Casimir5} follows trivially  once~\eno{calWCasimir} is established.

To show~\eno{calWCasimir}, it is convenient to appeal to the embedding space formalism by passing to the $(n+2)$-dimensional Minkowski space, $\mathbb{R}^{n+1,1}$. The AdS boundary coordinates are identified with coordinates $P \in \mathbb{R}^{n+1,1}$ on the null cone, satisfying $P^2\equiv P\cdot P=0$, while  Euclidean AdS$_{n+1}$ is given by the hyperboloid $Z^2=-1$. 
The bulk-to-bulk and bulk-to-boundary propagators take the form
 \eqn{GKEmbedding}{
 \hat{G}_{\Delta}(W,Z) = \left({\xi(W,Z) \over 2} \right)^{\Delta} \pFq{2}{1}{{\Delta \over 2},{\Delta+1 \over 2}}{\Delta-n/2+1}{\xi(W\!,\!\!Z)^2} \qquad 
 \hat{K}_\Delta(P,Z) = {1 \over (-2 P \cdot Z)^\Delta}\,,
 }
where $Z^2=W^2=-1,P^2=0$, and
\eqn{xiEmbedding}{
\xi(W,Z) = {1 \over (-W\cdot Z)}\,.
}

In embedding space, the conformal group $SO(n+1,1)$ acts linearly as the Lorentz group of $\mathbb{R}^{n+1,1}$. The generators ${\cal L}$  take the form of Lorentz generators in $n+2$ dimensions,
\eqn{LAB}{
{\cal L}^{(X)}_{AB} = -i\left(X_A {\partial \over \partial X^B} - X_B {\partial \over \partial X^A}\right),
}
and thus the Casimir acts as 
\eqn{}{
\left( {\cal L}^{(Z)} \right)^2 \phi(Z) = \left(-Z^2 \partial_Z^2 + Z \cdot  \partial_Z(n+Z\cdot \partial_Z) \right) \phi(Z) = \nabla^2_{\rm AdS} \phi(Z)
}
on arbitrary scalar functions on the AdS hypersurface $Z^2=-1$. For instance, 
\eqn{KGCasimir}{ 
\left( {\cal L}^{(Z)} \right)^2 \hat{K}_{\Delta}(P,Z) = m_\Delta^2 \hat{K}_\Delta(P,Z) \qquad \left( {\cal L}^{(Z)} \right)^2 \hat{G}_{\Delta}(W,Z) = m_\Delta^2 \hat{G}_\Delta(W,Z) \quad W \neq Z\,,
}
which follows from the equations of motion of the propagators. To demonstrate~\eno{calWCasimir} we will also need the following identities, which can be easily checked starting from~\eno{LAB}:
\eqn{UsefulAction1}{
\left({\cal L}^{(Z)}\right)^2 \left({\xi(W,Z) \over 2}\right)^\Delta &= m_\Delta^2 \left({\xi(W,Z) \over 2}\right)^\Delta - 4\Delta(\Delta+1) \left({\xi(W,Z) \over 2}\right)^{\Delta+2} \,,
}
and
\eqn{UsefulAction2}{
\left( {\cal L}_{AB}^{(Z)} \hat{K}_{\Delta_1}(P,Z) \right) \left({\cal L}^{(Z)}{}^{AB} \left({\xi(W,Z) \over 2}\right)^{\Delta_2} \right) &= 2\Delta_1\Delta_2 \hat{K}_{\Delta_1}(P,Z) \left({\xi(W,Z) \over 2}\right)^{\Delta_2} \cr 
 &- 4\Delta_1 \Delta_2 \hat{K}_{\Delta_1+1}(P,Z) \hat{K}_{-1}(P,W) \left({\xi(W,Z) \over 2}\right)^{\Delta_2+1},
}
where $\hat{K}_{-1}(P,W)$ should be understood formally as $(-2P \cdot W)$.

Next, we work out the action of the conformal Casimir $\left({\cal L}^{(1)} + {\cal L}^{(2)} + {\cal L}^{(3)}\right)^2$ on the generic geodesic integral defined in~\eno{BasicDiag}, where now the superscripts $(1),(2)$ and $(3)$ stand for generators built from the null coordinates $P_1, P_2$ and $P_3$ respectively. Following the technique of Refs.~\cite{DHoker:1999mqo,Hijano:2015zsa}, we isolate the part of~\eno{BasicDiag} which depends on $P_1,P_2$ and $P_3$:
\eqn{Fdef}{
\hspace{-0.2cm} F_{\Delta_L,\Delta_C,\Delta_R} (P_1,P_2,P_3,Z^\prime) \equiv \int_{Z\in \gamma_{12}}\!\!\!\!\! \hat{K}_{\Delta_1}(P_1,Z) \hat{K}_{\Delta_2}(P_2,Z) \hat{K}_{\Delta_L}(P_3,Z) \left({\xi(Z,Z^\prime) \over 2}\right)^{\Delta_C}\!\! \hat{K}_{\Delta_R}(P_3,Z^\prime)\,,
}
and use its invariance under simultaneous $SO(n+1,1)$ rotations of $P_1,P_2,P_3$ and $Z^\prime$ to conclude
\eqn{Finvariance}{
\left( {\cal L}^{(1)}_{AB} +  {\cal L}^{(2)}_{AB}+ {\cal L}^{(3)}_{AB} \right) F_{\Delta_L,\Delta_C,\Delta_R}(P_1,P_2,P_3,Z^\prime) = -{\cal L}^{(Z^\prime)}_{AB} F_{\Delta_L,\Delta_C,\Delta_R}(P_1,P_2,P_3,Z^\prime)\,,
}
which immediately leads to
\eqn{FCasimir}{
\left( {\cal L}^{(1)} +  {\cal L}^{(2)} + {\cal L}^{(3)} \right)^2 F_{\Delta_L,\Delta_C,\Delta_R}(P_1,P_2,P_3,Z^\prime) = \left( {\cal L}^{(Z^\prime)} \right)^2 F_{\Delta_L,\Delta_C,\Delta_R}(P_1,P_2,P_3,Z^\prime)\,.
}
At this stage, the proof by conformal Casimir equation employed in the case of the \emph{four-point} geodesic bulk diagram~\cite{Hijano:2015zsa} is essentially complete, since the $Z^\prime$ dependence on the analogously defined scalar function $F$ is contained entirely in a bulk-to-bulk propagator, so that one can use~\eno{KGCasimir} to immediately conclude that the appropriately defined $F$ (and as a consequence the full four-point geodesic bulk diagram) is an eigenfunction of the quadratic Casimir with the correct eigenvalue. 
However, for the five-point geodesic bulk diagram, the $Z^\prime$ dependence in $F$ enters via a product of a bulk-to-boundary propagator and a factor of $\xi(Z,Z^\prime)^\Delta$, such that $F$ in~\eno{Fdef} is \emph{not} an eigenfunction of the conformal Casimir. (This in turn is the reason the holographic dual is a sum over geodesic bulk diagrams, rather than a single geodesic bulk diagram --- the combination of geodesic bulk diagrams turns out to be an eigenfunction even though no individual diagram is one.)
Using the chain rule and \eno{KGCasimir}-\eno{UsefulAction2}, we obtain instead (suppressing the arguments of $F$)
\eqn{FCasimir2}{
\left( {\cal L}^{(Z^\prime)} \right)^2 F_{\Delta_L,\Delta_C,\Delta_R} &= m_{\Delta_C+\Delta_R}^2 F_{\Delta_L,\Delta_C,\Delta_R} - 4\Delta_C(\Delta_C+1) F_{\Delta_L,\Delta_C+2,\Delta_R} \cr 
&- 4\Delta_C \Delta_R F_{\Delta_L-1,\Delta_C+1,\Delta_R+1}\,.
}
From this we obtain the action of the conformal Casimir on the generic geodesic integral~\eno{BasicDiag}:
\eqn{BasicDiagCasimir}{
& \left( {\cal L}^{(1)} +  {\cal L}^{(2)} + {\cal L}^{(3)} \right)^2 \musepic{\figcalWdef} = m^2_{\Delta_C+\Delta_R} \musepic{\figcalWdef} \cr 
 &- 4\Delta_C(\Delta_C+1) \musepic{\figcalWdefC} 
 - 4\Delta_C \Delta_R \musepic{\figcalWdefLCR}.
}
We are now ready to evaluate the action of the conformal Casimir on the ``geodesic bulk diagram'' ${\cal W}_{\Delta_a,\Delta_b}(x_i)$ in~\eno{calW}. Our claim is that the weighted sum over geodesic diagrams~\eno{calW} \emph{is} in fact an eigenfunction of the Casimir with eigenvalue $C_2(\Delta_b)=m_{\Delta_b}^2$ as we now demonstrate. Acting with the quadratic Casimir as shown in the left hand side of~\eno{calWCasimir} and using~\eno{BasicDiagCasimir}, we obtain, analogous to~\eno{BasicDiagCasimir}, a combination of three terms, each of which is a double-sum over dummy integers $k_1, k_2$. By a few simple integer shifts of the dummy variables, we can bring each term to the same geodesic integral form, producing,
\eqn{calWCasimir2}{
  \left( {\cal L}^{(1)} +  {\cal L}^{(2)} + {\cal L}^{(3)} \right)^2 \!\!\! \sum_{k_1,k_2=0}^\infty\!\!\! c_{k_1,k_2} \!\!\!\!\!\!\!\! \musepic{\figcalW} \!\!\!\! = \!\! \sum_{k_1,k_2=0}^\infty \!\!\! \widetilde{c}_{k_1,k_2}\!\!\!\!\!\!\! \musepic{\figcalW},
}
where we have
\eqn{ctilde}{
\widetilde{c}_{k_1,k_2} &\equiv \bigg( m_{\Delta_b+2k_1}^2\: c_{k_1,k_2} -4 (\Delta_{ab,3}+k_1+k_2-2)(\Delta_{ab,3}+k_1+k_2-1)\: c_{k_1-1,k_2-1} \cr 
  &\qquad -4(\Delta_{b3,a}+k_1-k_2-1)(\Delta_{ab,3}+k_1+k_2-1)\: c_{k_1-1,k_2} \bigg)\,.
}
So to prove~\eno{calWCasimir} all that remains to be shown is that 
\eqn{ctildeToShow}{
\widetilde{c}_{k_1,k_2} \stackrel{!}{=} m_{\Delta_b}^2 c_{k_1,k_2} \qquad \forall\: k_1,k_2 = 0,1,2,\ldots\,.
}
Substituting the  $c_{k_1,k_2}$ coefficients from~\eno{cDef}  in the combination $\widetilde{c}_{k_1,k_2}  -  m_{\Delta_b}^2 c_{k_1,k_2}$, we find that the demonstration of~\eno{ctildeToShow} hinges on showing the following non-trivial identity between hypergeometric ${}_3F_2$ functions:
\eqn{3F2ToShow}{
{\textstyle   E(D-B)\: \pFq{3}{2}{A+1,B,C}{D+1,E}{1} + B(E-C)\: \pFq{3}{2}{A+1,B+1,C}{D+1,E+1}{1} - D E\: \pFq{3}{2}{A,B,C}{D,E}{1} \stackrel{!}{=} 0,}
}
where
\eqn{ABCDEF}{
\centering
A = -k_1 \qquad B=-k_2 \qquad C = \Delta_{ab3,}-n/2 \cr 
D = -k_1 + \Delta_{a3,b} \qquad E=-k_2+\Delta_{b3,a}\,.
}
Indeed with the help of hypergeometric identities~\cite{wolfram}
\eqn{3F2Id1}{
{\textstyle D E (D + 1) \: \left( \pFq{3}{2}{A,B,C}{D,E}{z} - \pFq{3}{2}{A+1,B,C}{D+1,E}{z} \right) + z  B C (D - A)\: \pFq{3}{2}{A+1,B+1,C+1}{D+2,E+1}{z} =0 \,,}
}
and~\cite[Eqn.~3.7.14]{andrews1999special}
\eqn{3F2Id2}{
{\textstyle E \: \pFq{3}{2}{A,B,C}{D,E}{1} -(E-A)\: \pFq{3}{2}{A,B+1,C+1}{D+1,E+1}{1} - {A (D-B)(D-C) \over D(D+1)}\: \pFq{3}{2}{A+1,B+1,C+1}{D+2,E+1}{1} = 0\,,}
}
 we conclude that the left hand side of~\eno{3F2ToShow} vanishes identically. 
 
 This completes the proof of~\eno{calWconfwave}, that the geodesic bulk diagram~\eno{calW} (and thus~\eno{calWalt}) is proportional to the five-point global conformal block with external scalars.

\section{Discussion}
\label{DISCUSSION}

In this paper, we have proposed the holographic object which computes the global five-point conformal block (for external scalar operators) in any spacetime dimension. It is given by a sum over geodesic bulk diagrams as described in~\eno{calWalt}-\eno{BasicDiagG} (equivalently by~\eno{calW}-\eno{cDef}).
Explicit expressions for the global five-point conformal block are already known in the literature,  obtained via the monodromy method~\cite{Alkalaev:2015fbw} and the shadow formalism~\cite{Rosenhaus:2018zqn}, so it is natural to ask how our results line up against previous results.
While we have provided conclusive proof by Casimir equation that our proposed five-point geodesic bulk diagram is the holographic object which computes the precise block (as explained in~\eno{calWconfwave}), we haven't  attempted an explicit match of our results against those of Refs.~\cite{Alkalaev:2015fbw,Rosenhaus:2018zqn}. However, we expect them to agree. 
We do note that the  coefficients~\eno{c3Def} appearing in the geodesic diagram~\eno{calWalt} look curiously similar to the analytic expression for the five-point global conformal block in arbitrary dimension computed in Ref.~\cite{Rosenhaus:2018zqn}, and in fact, a factor of the hypergeometric ${}_3F_2$ function with similar arguments appears in both places. Moreover, we have performed numerical checks for $n=1,2$ to confirm that the holographic dual of the conformal block proposed in this paper does indeed reproduce the low dimensional five-point conformal blocks of Ref.~\cite{Rosenhaus:2018zqn} to the numerical precision of choice.\footnote{We note that in Ref.~\cite{Rosenhaus:2018zqn}, what we call a conformal partial wave is referred to as a conformal block, and what we call a conformal block is referred to as a bare conformal block.}
 
 It would be interesting to generalize the geodesic bulk diagram of this paper to the setting of blocks with external spins and symmetric-traceless exchanges in the intermediate channels, as has been possible for the case of the four-point block~\cite{Hijano:2015zsa,Nishida:2016vds,Castro:2017hpx,Dyer:2017zef,Tamaoka:2017jce,Nishida:2018opl}, as well as extend the methods of this paper to obtain holographic duals of higher-point blocks. 
 For six-point blocks and higher, conformal blocks exist in multiple independent channels. Ref.~\cite{Rosenhaus:2018zqn} computed arbitrary-point global conformal blocks in the \emph{comb channel} in dimensions $n=1,2$, and they were found to be expressible in terms of generalized hypergeometric functions. The method presented here, however, does not rely on simplifications associated with low dimensional CFTs. Thus a natural next step is to leverage the strengths of this method to obtain higher-point geodesic bulk diagrams in the comb channel in arbitrary dimensions, as well as global conformal blocks in other channels, which have remained inaccessible so far. 
 The higher-point geodesic bulk diagrams may also be useful in the setup of a potentially higher-point conformal bootstrap~\cite{Rosenhaus:2018zqn,Rattazzi:2008pe,Caracciolo:2009bx,Poland:2011ey,Rychkov:2011et,ElShowk:2012ht,Simmons-Duffin:2016gjk}.
  It would also be interesting to investigate the connection between the higher-point geodesic bulk diagrams, such as the one presented in this paper, and the holographic interpretation of higher-point large-$c$ conformal blocks in AdS$_3$/CFT$_2$ as Steiner trees~\cite{Alkalaev:2018nik}.
  
The construction of the holographic dual of the five-point block relied on the knowledge of a novel propagator identity --- one involving a bulk integral over a product of two bulk-to-bulk propagators and a bulk-to-boundary propagator~\cite{JP:inprep}. 
This identity (and its generalizations) further enables  the conformal block decomposition of five-point bulk diagrams, as well as the direct evaluation of various AdS one-loop diagrams in position space~\cite{JP:inprep}. It would be interesting to see how these propagator identities supplement the discussion of higher-loop diagrammatics in AdS~\cite{Penedones:2010ue,Fitzpatrick:2011dm,Aharony:2016dwx,Cardona:2017tsw,Giombi:2017hpr,Yuan:2017vgp,Yuan:2018qva,Bertan:2018khc,Liu:2018jhs,Ghosh:2018bgd}.
  
 As demonstrated in section~\ref{INTRO}, the $p$-adic formulation was particularly useful in informing the discussion of the hitherto unknown five-point geodesic diagram over reals, and we expect this to continue to hold true for higher-point geodesic diagrams as well~\cite{JP:inprep}. 
 In fact, the lack of descendants in $p$-adic CFTs is the main simplification which allowed us to do the $p$-adic computations fairly easily and  grants the $p$-adic framework its practical utility as a powerful computational device. As commonly witnessed in $p$-adic holography (and seen above in the context of the five-point conformal partial wave), the $p$-adic answers essentially capture the information regarding the leading primary contribution in the real framework;  the derivative contribution can in principle be subsequently fixed from conformal symmetry.
 We hope to apply the unique power of the $p$-adic setup to other settings in the future.

\subsection*{Acknowledgments}
I thank C.B.\ Jepsen for valuable discussions and extensive collaboration during the early phases of this work, for helpful comments on an earlier version of the draft, and for collaboration on related projects.

\appendix
\section{Propagator identities}
\label{PROP}

Two propagator identities  mentioned in section~\ref{INTRO} are~\cite{Hijano:2015zsa}:
\eqn{KKexpansion}{
\hat{K}_{\Delta_1}(x_1,z) \hat{K}_{\Delta_2}(x_2,z)  &= \sum_{k=0}^\infty  {2 \over \beta_\infty(2\Delta_1+2k,2\Delta_2+2k)}  {(-1)^k \over k!} {(\Delta_1)_k (\Delta_2)_k \over (2\Delta_{12,}+k-n/2)_k} \cr 
 & \qquad \times \int_{w\in \gamma_{12}} \hat{K}_{\Delta_{1}}(x_1,w) \hat{K}_{\Delta_{2}}(x_2,w)  \hat{G}_{2\Delta_{12,}+2k}(z,w)\cr 
 &=  {2 \over \beta_\infty(2\Delta_1,2\Delta_2)} \int_{w\in \gamma_{12}}  \hat{K}_{\Delta_{1}}(x_1,w) \hat{K}_{\Delta_{2}}(x_2,w)  \left({\xi(z,w) \over 2}\right)^{2\Delta_{12,}}\,,
}
where we made use of~\eno{G2xi} to obtain the final equality, and
\eqn{GG}{
\int_{z \in {\rm AdS}} \hat{G}_{\Delta_a}(a,z) \hat{G}_{\Delta_b}(b,z) = { {1\over N_{\Delta_b}} \hat{G}_{\Delta_a}(a,b) - {1\over N_{\Delta_a}} \hat{G}_{\Delta_b}(a,b)  \over m_{\Delta_a}^2-m_{\Delta_b}^2}\,,
}
where $N_\Delta$ was defined in~\eno{GhatEOM}.

 \vspace{1em}
A useful, new propagator identity we made use of in section~\ref{OPE} is:
\eqn{KKGid}{
\int_{w\in \gamma_{12}} \!\!\!\!\! \hat{K}_{\Delta_1}(x_1,w) \hat{K}_{\Delta_2}(x_2,w) \hat{G}_{\Delta_a}(z,w) = {1\over 2}\beta_{\infty}(2\Delta_{a1,2},2\Delta_{a2,1})
&\sum_{k=0}^\infty {(\Delta_{a1,2})_k (\Delta_{a2,1})_k \over k! (\Delta_a-n/2+1)_k} \cr 
&\times  {\hat{K}_{\Delta_{a1,2}+k}(x_1,z) \hat{K}_{\Delta_{a2,1}+k}(x_2,z) \over (x_{12}^2)^{\Delta_{12,a}-k}}\,,
}
where the bulk geodesic $\gamma_{12}$ joins boundary points $x_1, x_2 \in \partial$AdS$_{n+1}, z\in$AdS$_{n+1}$.
To prove this, we just need to show that
\eqn{KKxiId}{
\int_{w\in \gamma_{12}} \!\!\!\!\! \hat{K}_{\Delta_1}(x_1,w) \hat{K}_{\Delta_2}(x_2,w)\! \left({\xi(z,w) \over 2}\right)^{\Delta_0}\! = {1\over 2}\beta_{\infty}(2\Delta_{01,2},2\Delta_{02,1})
 {\hat{K}_{\Delta_{01,2}}(x_1,z) \hat{K}_{\Delta_{02,1}}(x_2,z) \over (x_{12}^2)^{\Delta_{12,0}}}\,,
}
since setting $\Delta_0=\Delta_{a}+2k$, multiplying the left- and right-hand sides of~\eno{KKxiId} by a factor of ${1\over k!} {(\Delta_a)_{2k} \over (\Delta_a-n/2+1)_k}$, and  summing over all $k$, we recover~\eno{KKGid}.
Moreover, we note that~\eno{KKxiId} is more general than~\eno{KKexpansion}; thus proving~\eno{KKxiId} indirectly furnishes a proof for the inverse relation~\eno{G2xi} for arbitrary $n$.

 To prove~\eno{KKxiId}, use conformal transformations to put $x_1,x_2,z$ on a common hyperbolic plane (s.t.\ $z=(u_z,x_z) \in \mathbb{H}^2$ with $x_1,x_2$ located on $\partial\mathbb{H}^2$). Using the parameter $\theta \in [0,\pi]$ to describe the location on the semi-circular geodesic joining $x_1$ and $x_2$, the integral reduces to (assuming $x_1>x_2$),
\eqn{}{
{u_z^{\Delta_0} \over (x_{12}^2)^{\Delta_{12,0}}} 
\int_{0}^\pi {d\theta \over \sin \theta}  
{1 \over (\tan {\theta\over 2})^{\Delta_1} (\cot {\theta\over 2})^{\Delta_2}} 
\left({\sin \theta \over \lambda_1 + \lambda_2 \cos \theta} \right)^{\Delta_0}
}
with $\lambda_1 = 2u_z^2+ {1\over 2}( x_{12}^2 +(x_{1z}+x_{2z})^2)$ and $\lambda_2 = (x_{1z}+x_{2z})x_{12}$, which can be readily evaluated to give the right-hand side of~\eno{KKxiId}.

\bibliographystyle{ssg}
\bibliography{draft} 
\end{document}